\documentclass[%
reprint,
superscriptaddress,
 amsmath,amssymb,
prb,
]{revtex4-1}

\newcommand*{\degc}{$\!^\circ$C}

\usepackage{graphicx}
\usepackage{dcolumn}
\usepackage{bm}
\usepackage{makecell}
\usepackage{multirow}

\begin{document}

\preprint{APS/123-QED}

\title{Inhomogeneous ferromagnetism mimics signatures of \\the topological Hall effect in SrRuO$_3$ films}

\author{Gideok Kim}
\email{gideokkim@skku.edu}
\affiliation{Max-Planck-Institute for Solid State Research, Heisenbergstrasse 1, 70569 Stuttgart, Germany}%
\author{K. Son}%
\affiliation{Max-Planck-Institute for Intelligent systems, Heisenbergstrasse 1, 70569 Stuttgart, Germany}%
\author{Y. Eren Suyolcu}%
\affiliation{Department of Materials Science and Engineering, Cornell University, Ithaca, NY 14853, USA}%
\affiliation{Max-Planck-Institute for Solid State Research, Heisenbergstrasse 1, 70569 Stuttgart, Germany}
\author{L. Miao}%
\affiliation{Department of Physics, Laboratory of Atomic and Solid State Physics, Cornell University, Ithaca, NY 14853, USA}
\author{N. J. Schreiber}%
\affiliation{Department of Materials Science and Engineering, Cornell University, Ithaca, NY 14853, USA}
\author{H. P. Nair}%
\affiliation{Department of Materials Science and Engineering, Cornell University, Ithaca, NY 14853, USA}
\author{D. Putzky}
\affiliation{Max-Planck-Institute for Solid State Research, Heisenbergstrasse 1, 70569 Stuttgart, Germany}%
\author{M. Minola}%
\affiliation{Max-Planck-Institute for Solid State Research, Heisenbergstrasse 1, 70569 Stuttgart, Germany}%
\author{G. Christiani}%
\affiliation{Max-Planck-Institute for Solid State Research, Heisenbergstrasse 1, 70569 Stuttgart, Germany}%
\author{P. A. van Aken}%
\affiliation{Max-Planck-Institute for Solid State Research, Heisenbergstrasse 1, 70569 Stuttgart, Germany}%
\author{K. M. Shen}%
\affiliation{Department of Physics, Laboratory of Atomic and Solid State Physics, Cornell University, Ithaca, NY 14853, USA}
\affiliation{Kavli Institute at Cornell for Nanoscale science, Ithaca, NY 14853, USA}
\author{D. G. Schlom}%
\affiliation{Department of Materials Science and Engineering, Cornell University, Ithaca, NY 14853, USA}
\affiliation{Kavli Institute at Cornell for Nanoscale science, Ithaca, NY 14853, USA}
\author{G. Logvenov}%
\affiliation{Max-Planck-Institute for Solid State Research, Heisenbergstrasse 1, 70569 Stuttgart, Germany}%
\author{B. Keimer}%
\email{B.Keimer@fkf.mpg.de}
\affiliation{Max-Planck-Institute for Solid State Research, Heisenbergstrasse 1, 70569 Stuttgart, Germany}%

\date{\today}

\begin{abstract}
Topological transport phenomena in magnetic materials are a major topic of current condensed matter research. One of the most widely studied phenomena is the ``topological Hall effect'' (THE), which is generated via spin-orbit interactions between conduction electrons and topological spin textures such as skyrmions. We report a comprehensive set of Hall effect and magnetization measurements on epitaxial films of the prototypical ferromagnetic metal SrRuO$_3$ whose magnetic and transport properties were systematically modulated by varying the concentration of Ru vacancies. We observe Hall effect anomalies that closely resemble signatures of the THE, but a quantitative analysis demonstrates that they result from inhomogeneities in the ferromagnetic magnetization caused by a non-random distribution of Ru vacancies. As such inhomogeneities are difficult to avoid and are rarely characterized independently, our results call into question the identification of topological spin textures in numerous prior transport studies of quantum materials, heterostructures, and devices. Firm conclusions regarding the presence of such textures must meet stringent conditions such as probes that couple directly to the non-collinear magnetization on the atomic scale.
\end{abstract}

\maketitle

\section{Introduction}
Ferromagnetic and nearly ferromagnetic metals are archetypal platforms for the investigation of topological phenomena in quantum materials.  The “anomalous Hall effect” (AHE) in a metallic ferromagnet is proportional to the ferromagnetic magnetization, with a coefficient that depends on the Berry curvature in momentum space and thus contains information on the topological properties of its band structure \cite{Haldane2004,Nagaosa2010}.  More recently, an additional contribution (termed “topological Hall effect”, THE) was discovered in nearly ferromagnetic metals with superstructures of real-space topological defects, namely skyrmions \cite{Roessler2006,Neubauer2009}. This contribution arises when conduction electrons accumulate a Berry phase upon traversing the skyrmion lattice, and manifests itself as a sharp enhancement of the Hall signal as the skyrmion-lattice phase is entered by varying the temperature or the external field.

\begin{figure*}[t]
	\includegraphics[width=4.5 in]{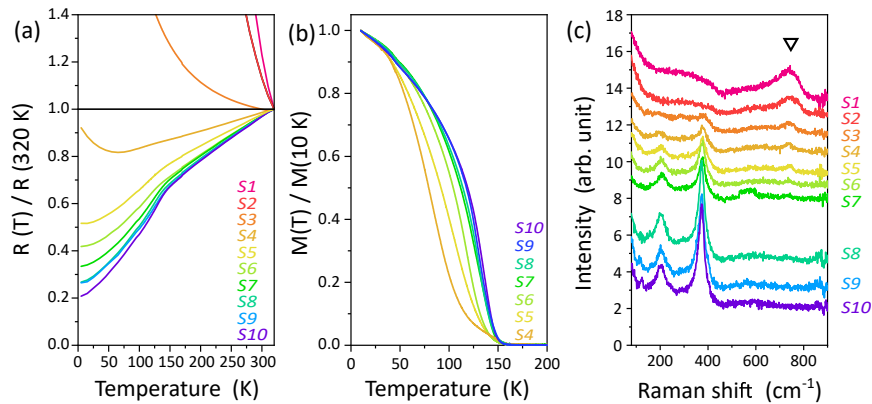}
	\caption{\label{fig:characterization}\textbf{Physical properties of SrRuO$_3$ films with varying Ru vacancy concentration.} (a) Normalized resistance curves. (b) Normalized field-cooled magnetization measurement at 0.01 T. Magnetization curves of Sample S1-3 are not plotted here due to the absence of ferromagnetic moments. The raw data without normalization are presented in the Supplemental Material. (c) Polarized Raman spectra with $z(XX)\bar{z}$ configuration. The triangle indicates the Raman-active mode induced by Ru vacancies.}
\end{figure*}

The AHE was initially discovered and explored in bulk ferromagnets, but has become a standard diagnostic of ferromagnetism in device structures where direct measurements of the magnetization are impractical. Likewise the THE, which was originally established in bulk compounds with skyrmion-lattice phases that had been extensively characterized by neutron diffraction, has recently been widely used as a fingerprint of skyrmions in thin films and heterostructures where direct measurements of the non-collinear magnetization could not be performed \cite{Matsuno2016,Ohuchi2018,Wang22019,Vistoli2019,Wang2018,Li2020,Qin2019,Meng2019,Gu2019,Sohn2020,Ziese2019,Zhang2020,Bartram2019}. Information about topological spin textures in such systems is important in view of the emerging use of skyrmions in data processing and storage devices (“skyrmionics”) \cite{Back2020}. However, the indirect identification of topological structures based on THE experiments has been called into question, because 
sharp enhancements of the Hall effect in specific regimes of temperatures and magnetic fields can also arise from different mechanisms, including electronic inhomogeneity due to variations in materials composition and superposition of Hall signals from different materials and interfaces in heterostructures \cite{Gerber2018,Kan2018,Kan2020}. 

With this motivation, we have carried out a comprehensive set of measurements of the Hall effect and magnetization in epitaxial films of a single model material, the ferromagnetic metal SrRuO$_3$ (SRO). In thin-film form, SRO is widely employed as a metallic electrode in oxide electronics and as a ferromagnetic component of spintronic devices \cite{Koster2012}. The AHE of bulk SRO is positive below its Curie temperature $T_{C} = 165$ K, but exhibits a sign reversal upon cooling below $T_S \approx 120$ K \cite{Koster2012}. This behavior can be attributed to the confluence of an intrinsic contribution from the Berry curvature of the band dispersions and an extrinsic contribution due to spin-orbit scattering from defects. Due to inversion-symmetry breaking at interfaces and/or interfacial exchange interactions, the magnetic structure of SRO can become non-collinear in thin-film structures  \cite{Kim2012}, and theoretical work has led to the prediction of topological defects and defect superstructures in this situation \cite{Koshibae2017,Mohanta2019}. Hall effect measurements both on SRO films and on heterostructures comprising SRO layers have indeed uncovered sharp enhancements of the Hall signal akin to those of classical skyrmion-lattice compounds, which were interpreted as evidence for skyrmions generated by Dzyalonshinskii-Moriya interactions at the substrate interface and/or competing exchange interactions at interfaces with other magnetic materials \cite{Matsuno2016,Ohuchi2018}. 

Recently, however, an alternative interpretation was proposed based on the superposition of different AHE signals due to spatial inhomogeneity \cite{Kan2018,Kan2020,Miao2020,Wysocki2020,Kimbell2020}. Whereas a model based on this hypothesis yielded a good description of transport data on ultrathin SRO films, a thorough assessment of this scenario requires information on the ferromagnetic magnetization which controls the magnitude of the AHE. We performed our measurements on thick films where interfacial effects are negligible and where the ferromagnetic order parameter and its inhomogeneity could be quantitatively characterized by magnetometry. An elementary model of the AHE based on this information yields a quantitative description of the Hall effect maxima purely based on inhomogeneous ferromagnetism.

\section{Experimental details}

Thin films were grown on (LaAlO$_3$)$_{0.3}$-(SrAl$_{0.5}$Ta$_{0.5}$O$_3$)$_{0.7}$ (LSAT) (001) single-crystalline substrates (CrysTec GmbH) using either a reactive sputtering system developed at the Max Planck Institute for Solid State Research or oxide MBE. 
For reactive sputtering, argon and oxygen gas were supplied via a mass flow controller. 
The pressures P$_{O_2}$ and P$_{total}$ were 50 and 100 mTorr, respectively. 
Substrates were glued with a platinum paste to pure nickel blocks and heated with an infrared laser. 
The substrate temperature was monitored using a radiative pyrometer using the emissivity of $\epsilon_\mathrm{LSAT}$=0.92. 
The structural quality of the films was confirmed by high-resolution X-ray diffraction (XRD) with a Cu K-$\alpha$ source ($\lambda$ $\approx$ 1.5406 \r{A}) and by transmission electron microscopy. 
All the samples investigated in this study are listed in table \ref{tab:table1}. 
The growth parameters for the films grown via molecular beam epitaxy (MBE) have been presented elsewhere \cite{Nair2018-113}.

\begin{table}[h]
	\caption{\label{tab:table1}%
		List of samples grown by reactive sputtering. 
		The samples are listed in order of their residual resistivity ratios. A sample labeled as weakly metallic shows a resistivity minimum as a function of temperature. Sr214 and Ru stand for pressed polycrystalline Sr$_2$RuO$_4$ and Ru metal targets, respectively. The last column indicates whether the ground state is paramagnetic (PM) or ferromagnetic (FM).  
	}
	\begin{ruledtabular}
		\begin{tabular}{ccccc}
			\textrm{Sample}&
			\textrm{Transport}&
			\textrm{$T_\mathrm{growth}$ (\degc)}&
			\textrm{Target}&
			\textrm{Magn.}\\
			\colrule
			$S1$& Insulating & 440& Sr214 & PM \\
			$S2$& Insulating&440&  Sr214+Ru & PM  \\
			$S3$& Insulating &530&  Sr214 & PM  \\
			$S4$& Weakly metallic &600&  Sr214 & FM  \\
			$S5$& Metallic &530&  Sr214+Ru & FM  \\
			$S6$& Metallic &600&  Sr214+Ru & FM  \\
			$S7$& Metallic &700&  Sr214 & FM  \\
			$S8$& Metallic &700&  Sr214 + Ru & FM  \\
			$S9$& Metallic &770&  Sr214 & FM \\
			$S10$& Metallic &770&  Sr214+Ru & FM \\
		\end{tabular}
	\end{ruledtabular}
\end{table}

The Raman spectra were measured with a Jobin-Yvon LabRam HR800 spectrometer (Horiba Co.) combined with a dedicated confocal microscope. The short depth of focus allows measurements of films with thicknesses of $\approx$ 10 nm. The samples were illuminated with a He-Ne laser with wavelength 632.8 nm (red), and the scattered light was collected from the sample surface with a 100$\times$ long working distance objective lens. The experiments were performed in backscattering geometry with (\textit{a},\textit{b})-axis polarized light propagating along the crystallographic \textit{c}-axis, which is denoted as $z(XX)\bar{z}$ in Porto's notation.
SRO has a space group $Pnma$, which results in a lower symmetry than the simple perovskite structure ($Pm$\=3$m$).
Therefore SRO has 24 Raman-active phonon modes, 7$A_{g}$+5$B_{1g}$+7$B_{2g}$+5$B_{3g}$, according to the group theory analysis. 
The measurement was restricted by the experimental setup and the shape of thin film samples to the $z(XX)\bar{z}$ geometry, in which the propagation of light is parallel to the surface normal, and the polarization of the light is parallel to the Ru-O bonding direction. 
In this geometry we could study phonon modes with $A_{1g}+ B_{1g}$ symmetry.

The electric resistance and Hall measurements were carried out using a Physical Property Measurement System  (Quantum Design Inc.). To implement the van der Pauw geometry, Pt metallic contacts were sputtered on four corners of square shaped samples (5 mm $\times$ 5 mm).
SQUID magnetometry was used to measure the magnetic properties. 
The magnetization curves were measured using a  Magnetic Property Measurement System (MPMS, Quantum Design Inc.) in the VSM mode.
The first order reversal curves were measured using a MPMS in the DC mode.

\section{Results and discussion}

Due to the high volatility of ruthenium oxides, SRO films grown from stoichiometric targets contain Ru vacancies that are known to reduce the electrical conductivity and the Curie temperature \cite{Koster2012,Nair2018-113,Siemons2007,Dabrowski2004,Schraknepper2015,Boschker2019}. 
We modified the density of Ru vacancies and the physical properties by adding Ru metal to the target and by varying the growth temperature. Following recent work on Sr$_2$RuO$_4$ films \cite{Kim2019}, we monitored the intensity of a Raman-active mode attributable to Ru vacancies and found that it is inversely related to the residual resistivity ratio (RRR), $R_{xx} (T=320 K) / R_{xx} (T  \rightarrow 0)$, indicating that Ru vacancies are the major source of scattering of the conduction electrons (Fig. \ref{fig:characterization}(c)).

Figure \ref{fig:characterization}(a) shows the temperature ($T$) dependence of the longitudinal electrical resistance, $R_{xx}$, of the samples prepared in this way.  The RRR reflects the density of Ru vacancies. The samples were labeled $S1-10$, with higher numbers indicating higher RRRs. Samples $S1-3$ show semiconductor-like electrical transport with diverging $R_{xx}$ as $T \rightarrow 0$. Structural analysis shows that metallic samples with low and moderate Ru vacancy densities only exhibit the SRO phase, whereas electron microscopy data on insulating samples with higher vacancy content show possible evidence of secondary phase formation (see Supplement). 

In Figure \ref{fig:panorama}, we present the Hall resistance, $R_{xy}$, as a function of external magnetic field, $H$, for the metallic samples. $R_{xy}$ comprises two major contributions: the AHE, which is proportional to the ferromagnetic magnetization and thus follows its hysteresis loop \cite{Nagaosa2010}, and the ordinary Hall effect, which is responsible for the linear dependence of $R_{xy}(H)$ persisting above the coercive field $H_{C}$. The sign of the anomalous Hall resistance ($R_{xy,AHE}$) can be identified from the direction of jumps in the ascending branches of the hysteresis loops. In agreement with the literature, the AHE of nearly stoichiometric samples is positive (negative) for $T$ larger (smaller) than $T_{S} \approx 110$ K, respectively. Our synthesis method allowed us to vary $T_{S}$ over an exceptionally wide range (inset to Fig. \ref{fig:panorama}(e)). $R_{xy,AHE}$ and $R_{xx}$ of all samples (including the ones with severely degraded RRR and $T_S$) can be collapsed onto a single scaling plot (Fig. \ref{fig:panorama}(e)), which indicates the dominance of the intrinsic (extrinsic) contributions to the AHE at low (high) temperatures \cite{Haham2011,Haham2013}. In the metallic sample with the lowest RRR, $S4$, the impurity contribution with positive $R_{xy,AHE}$ prevails at all temperatures so that $R_{xy,AHE}$ does not change sign at all. 
\begin{figure*}[htb!]
	\includegraphics[width=7 in]{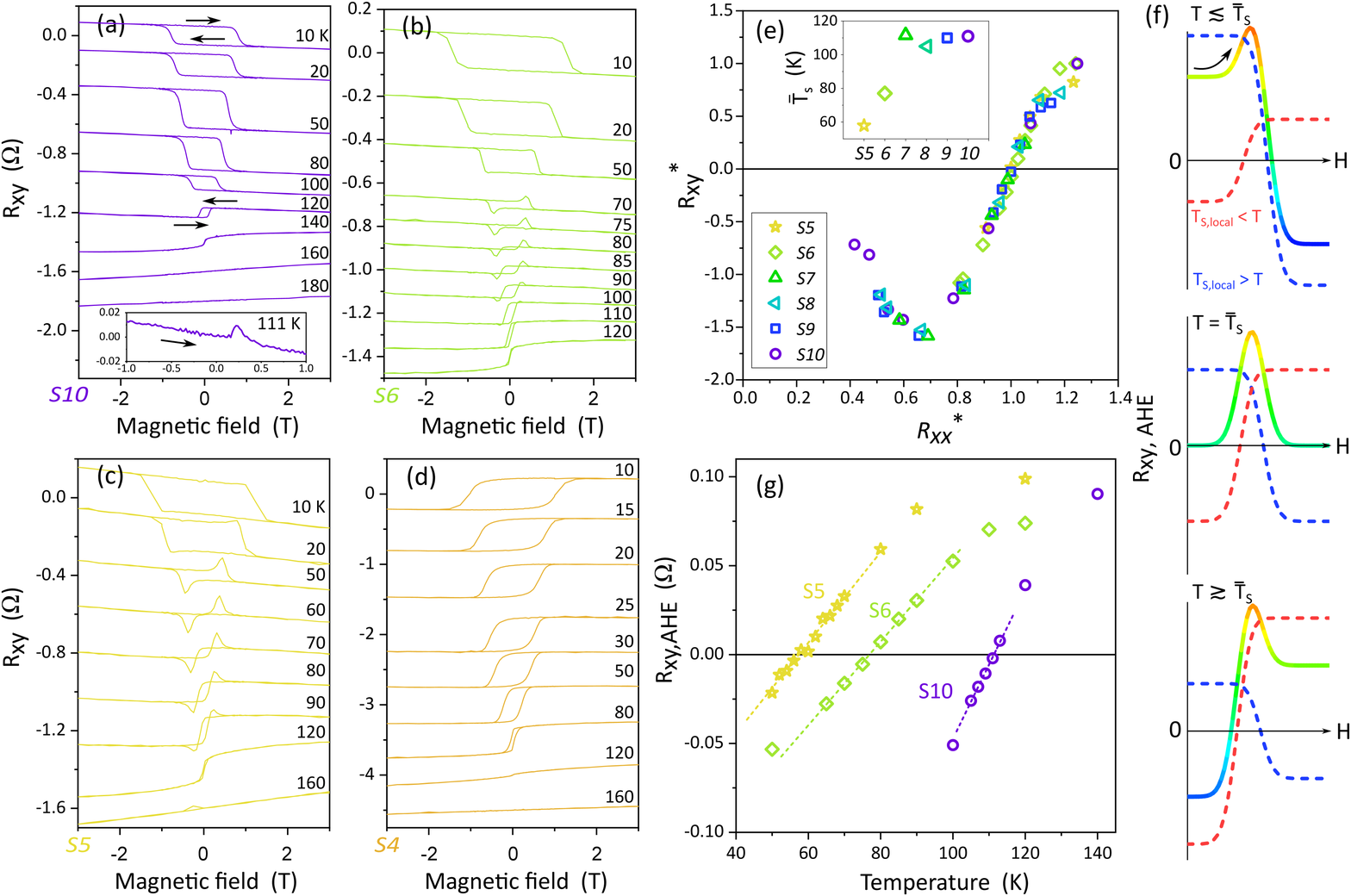}
	\caption{\label{fig:panorama} \textbf{Summary of Hall measurements.} (a)-(d) Hall resistance curves of four metallic samples. Curves are shifted in the $y$-direction for better visibility. The numbers in plots indicate the measurement temperatures in Kelvin. The black arrows indicate the direction of magnetic field ramp. The Hall measurements on the other samples are presented in the Supplemental Material. The inset to (a) shows the magnified view of the ascending branch. (e) Scaling behavior of the anomalous Hall resistance. The inset shows the sample dependence of $T_{S}$. Here, $R^{*}_{xy}$ is $R_{xy,AHE}(T)$ normalized by its maximum value, and $R^{*}_{xx}$ = $R_{xx}(T)$/$R_{xx}(T_S)$ \cite{Haham2011}. (g) Temperature dependence of $R_{xy,AHE}$. Dashed lines are the results of linear fit. (f) Schematic diagram describing the composite AHE that mimics the THE.}
\end{figure*}
The hysteresis loops of $R_{xy,AHE}$ in all other metallic samples (except $S4$) exhibit maxima for $H \approx H_C$ and $T \approx T_{S}$ (Fig. \ref{fig:panorama}(a)-(d)). Related anomalies of the Hall signal have previously been reported for SRO films, SRO heterostructures as well as heterostructures of different compounds, and have been widely attributed to the topological Hall effect \cite{Matsuno2016,Ohuchi2018,Wang2018,Qin2019,Meng2019,Li2020,Qin2019,Meng2019,Gu2019,Sohn2019,Ziese2019,Zhang2020,Bartram2019}. In our samples, however, the amplitude and temperature range of these maxima exhibit an extreme variation with defect concentration. In the most stoichiometric sample $S10$, the maxima are subtle and observable only over a range of a few K as shown in Fig. \ref{fig:panorama}(a), analogous to the THE in skymion-lattice compounds such as MnSi \cite{Neubauer2009}. In our sample $S5$ with a high Ru vacancy content, on the other hand, the maxima are much more pronounced and extend over a temperature range of more than 70 K. 

This finding is difficult to reconcile with the THE model and instead supports a scenario that attributes the Hall effect maxima to inhomogeneous ferromagnetism generated by unavoidable inhomogeneities in the distribution of Ru vacancies. In this scenario, the maxima arise for temperatures in the vicinity of the macroscopically averaged $\bar{T}_S$, when the current traverses regions with positive and negative $R_{xy,AHE}$ and different coercive fields. As an example, we consider the ascending branch of the hysteresis curve at a temperature $T \gtrsim \bar{T}_S$  (Fig. \ref{fig:panorama}(f)). In low magnetic fields, the saturated $M$ is antiparallel to the external field, and the overall $R_{xy,AHE}$ signal is reduced through a partial cancellation of positive and negative contributions from regions with $T_S < T  $ and $T_S > T  $, respectively. Regions with lower $T_S$ have higher defect densities and correspondingly lower $H_C$. As $H$ increases, the magnetization of these regions is reversed when the local $H_C$ is exceeded, while the magnetization of more stoichiometric domains with higher $H_C$ and negative $R_{xy,AHE}$ remains antiparallel to $H$. In this way, the partial cancellation is lifted and the overall $R_{xy,AHE}$ is enhanced. When $H$ exceeds $H_C$ of the more stoichiometric regions, the macroscopically averaged AHE signal decreases again \cite{Kan2018}. 

\begin{figure*}[t]
	\includegraphics[width=6.5 in]{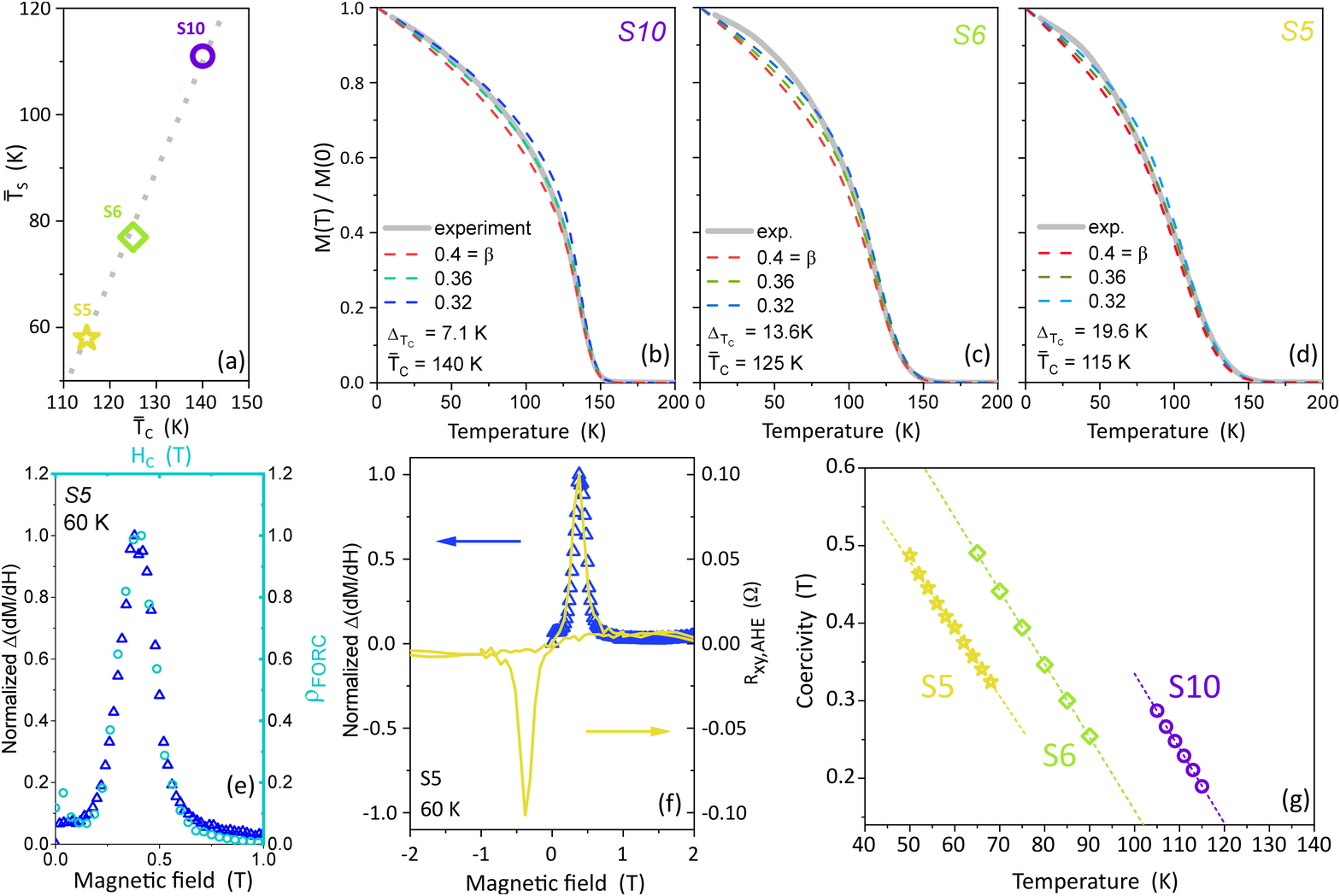}
	\caption{\label{fig:linear}\textbf{ Summary of magnetometry measurements.} (a) Correlation between $T_S$ and $T_C$. The line is a guide to the eye. (b)-(d) $M(T,H\rightarrow0)$ and modeled curves. The curves were computed using Gaussian averages with the parameters in the legends (see text). $\beta$ is the critical exponent in the power law, $(1-T/T_C)^\beta$.
	(e) Comparison between the FORC distribution $\rho_{FORC}$ and the derivative of the magnetic hysteresis curve of sample S5 at $T = 60$ K. Both were normalized to unity for comparison. (f) Comparison between the distribution of $H_C$ and the maximum in AHE. (g) Coercivity obtained from Gaussian fits the magnetometry measurements. Dashed lines are results of linear fits.}
\end{figure*}

At the heart of this picture is the inhomogeneous distribution of the ferromagnetic order parameter and coercive field, which was not available in prior work because the magnetization of the thin SRO layers could not be separated from the diamagnetic contribution of the substrates. Indirect information about these quantities was obtained from the longitudinal magnetoresistance \cite{Kan2018}. The $R_{xx}(T)$ traces of SRO indeed exhibit kinks at the ferromagnetic transition temperature $T_C$, and signatures of inhomogeneous broadening are observed in samples with lower RRR (Fig. \ref{fig:characterization}(a)), but the degree of broadening is difficult to quantify because the strength of the kinks also decreases with increasing RRR. In the presence of external magnetic fields, these difficulties are further compounded by incomplete knowledge of the mechanisms underlying the longitudinal magnetoresistance and its relation to the magnetization.
Moreover, since the films investigated in previous experiments were only 3-5 nm thick and the thickness was used to modulate the degree of inhomogeneity, the influence of the surface and the substrate interface could not be assessed. 

To address this challenge, we grew our films with thicknesses of 25-30 nm to enable accurate measurements of the ferromagnetic magnetization $M(T,H)$ -- a thermodynamic quantity that contains all of the information required for a definitive test of the inhomogeneity model. The Curie temperature $T_C$ measured by magnetometry decreases with increasing Ru vacancy content, in lockstep with the temperature at which the AHE changes sign (Fig. \ref{fig:linear}(a)).  This indicates that scattering of conduction electrons from defects introduced by Ru vacancies is the major source of degradation of ferromagnetic order, as expected for an itinerant-electron ferromagnet such as SRO. The $T$-dependent ferromagnetic order parameter shows unmistakable manifestations of rounding due to an inhomogeneous spread of Curie temperatures (Fig. \ref{fig:characterization}(b)). The degree of broadening can be accurately quantified by fitting $M(T, H \rightarrow 0)$ to a power law folded with a Gaussian $T_C$-distribution, and the results are presented in Fig. \ref{fig:linear}(b)-(d). (As expected, deviations from the power law are apparent at low $T$, but choosing different functional forms of $M(T)$ does not change the result significantly.) The width of this distribution, $\Delta T_C$, increases rapidly with defect concentration, in close analogy to the temperature range of the Hall effect maxima and in qualitative agreement with the inhomogeneity scenario for the Hall effect maxima.

\begin{figure*}[htb!]
	\includegraphics[width=6.5 in]{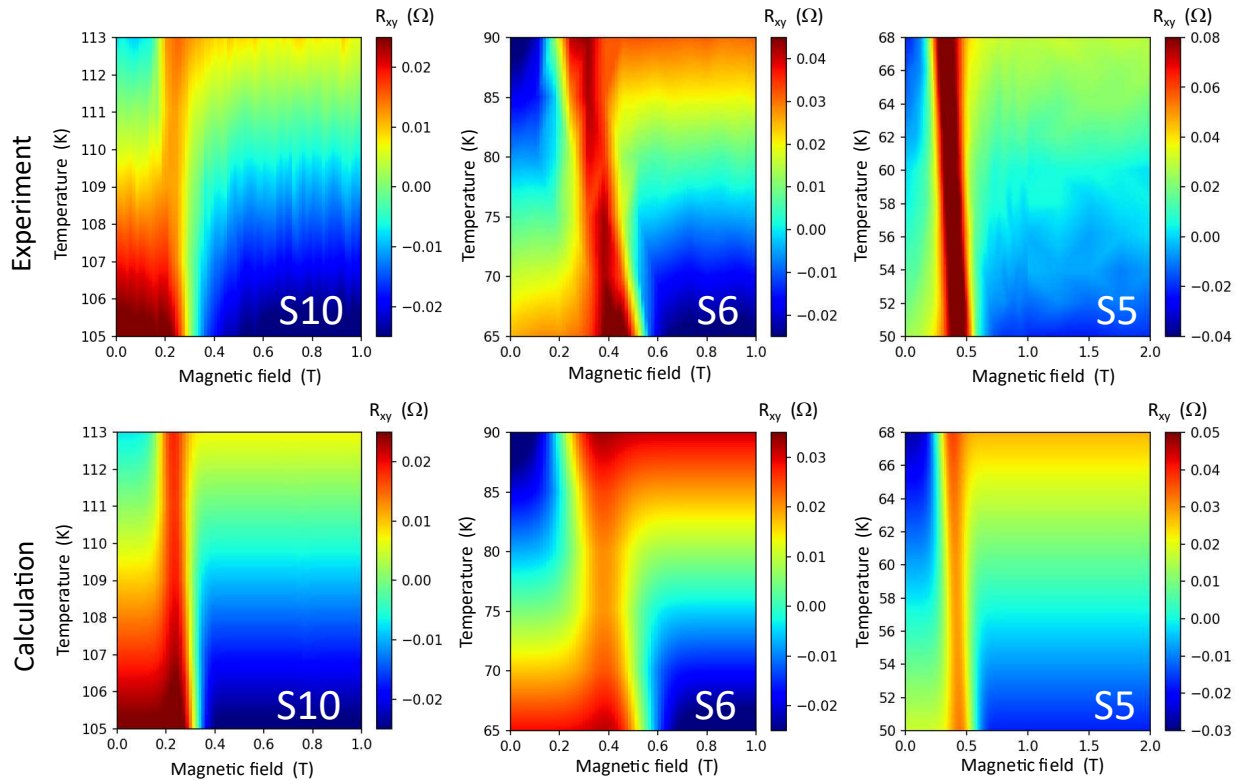}
	\caption{\label{fig:result}\textbf{Impact of inhomogeneity on the Anomalous Hall Effect.} Comparison between the experimental and calculated anomalous Hall resistance curves in the ascending branches of the hysteresis loops of samples S10, S6, and S5.}
\end{figure*}

According to this picture, the magnetic-field range of the Hall effect maxima is determined by the distribution of coercive fields, $\Delta H_C$, across a given sample (see supplement). Following standard practice, we have extracted $\Delta H_C$ from minor hysteresis loops with varying reversal fields. Figure \ref{fig:linear}(e) shows the outcome of this ``first-order reversal curve (FORC)'' analysis for a representative sample, along with the field dervative of $M(H)$. The quantitative agreement between both profiles (Fig. \ref{fig:linear}(e)) confirms that the broadening of the coercive fields indeed originates from an inhomogeneous $H_C$ distribution.
In Fig. \ref{fig:linear}(f) we also compare the width of this distribution to the width of the Hall effect maxima measured on the same sample at the same temperature. The excellent agreement provides further strong support for the inhomogeneity scenario.

The comprehensive set of magnetization data enabled us to devise a quantitative phenomenological model for the Hall effect maxima based purely on the conventional understanding of the AHE. Whereas we have obtained most of the essential information from our $M(T,H)$ measurements, the distribution of $T_S$ is a transport quantity that cannot be directly extracted from the magnetization measurements. However, our observation that scattering from Ru vacancies depresses the macroscopically averaged $\bar{T}_S$ and $\bar{T}_C$ at nearly identical rates implies that $\Delta T_S \propto \Delta T_C$ (Fig. \ref{fig:linear}(a)).

We averaged the AHE over regions with different $T_S$ and $H_C$ as follows: 
\begin{equation*}
R_{xy,AHE}(H;T) = \langle A(T_S;T) \cdot M (H, H_{C};T) \rangle_{T_{S}, H_{C}} = 
\end{equation*}
\begin{equation*}
\int\int\alpha_{A}(T-T_{S}){[2F_{step}(H-H_{C})-1]}\;g(T_S, H_C)\;dT_S\;dH_{C}
\end{equation*}

where $F_{step}$ is the Heaviside step function, $g$ is the bivariate Gaussian function
\begin{equation*}
	\begin{aligned}
		g(T_{S}, H_{C})=\frac{1}{2\pi\Delta_{T_S}\Delta_{H_{C}}\sqrt{ (1-\rho^2)}}
		\;exp[-\frac{1}{2(1-\rho^2)}\times\\(\frac{(T_{S}-\bar{T_{S}})^2}{\Delta_{T_{S}}^2} + \frac{(H_{C}-\bar{H_{C}})^2}{\Delta_{H_{C}}^2}-\frac{2\rho(T_S-\bar{T_S})(H_{C}-\bar{H_{C}})}{\Delta_{T_S}\Delta_{H_C}})]
	\end{aligned}
\end{equation*}
and the correlation coefficient $\rho$ is fixed to 0.75 in all calculations due to the strong correlation between $H_{C}$ and $T_{S}$. All coefficients used here are summarized in the Supplemental Material. 
In computing these averages, we take advantage of the observations that $\bar{A} = \alpha_{A}(T-\bar{T}_S$) (Fig. \ref{fig:panorama}(g)) and $\bar{H}_C = \alpha_{C}T-\beta_{C}$ (Fig. \ref{fig:linear}(g)) in the relevant range of temperatures, with coefficients that vary only weakly between samples with comparable $\bar{T}_C$, and make the straightforward assumption that these relations also hold for regions with different $T_S$ and $T_C$ traversed by the Hall current in any given sample.
Figure \ref{fig:result} shows the outcome of this analysis for three representative samples with widely varying amplitude and temperature range of the Hall effect maxima. The agreement between the measured and calculated $R_{xy,AHE}$ is excellent, which is remarkable because all parameters were fixed by independent magnetization measurements. 

\begin{figure}[htb!]
	\includegraphics[width=3.4 in]{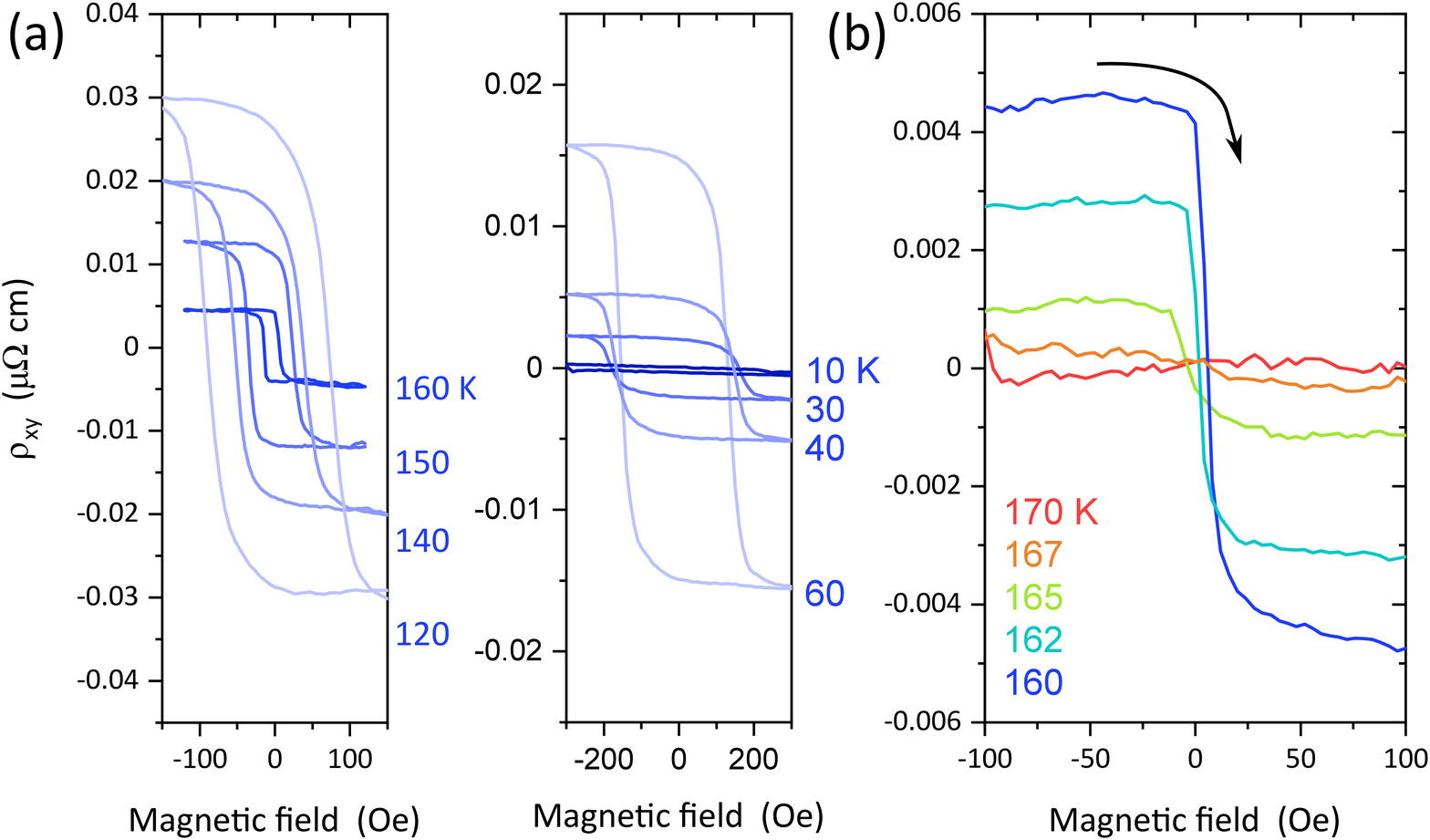}
	\caption{\label{fig:cornell} \textbf{Absence of the sign changing in the MBE-grown sample.} (a) Hall resistivity curves that comprise only negative AHE. (b) Detailed investigation of ascending branches of Hall resistivity curves measured near $T_C$.}
\end{figure}

Interestingly $S4$, the most inhomogeneous among the ferromagnetic metallic samples, does not exhibit Hall effect maxima (Fig. \ref{fig:panorama}(d)) due to the absence of sign reversal.
Further evidence of the importance of the sign-reversal can be found in the MBE-grown SRO thin film on a DyScO$_3$ (110) substrate with a remarkably high RRR of 130 owing to the perfect lattice match between SRO and DyScO$_3$ and to the adsorption-controlled growth\cite{Nair2018-113,Koster2012}.
Surprisingly, the extremely low concentration of defects keeps the sign of AHE negative at all temperatures in contrast to $S4$ that shows only positive AHE (see Fig. \ref{fig:cornell}(a)).
We closely investigated the ascending branches of the Hall resistivity curves near $T_C$ (Fig. \ref{fig:cornell}(b)), and we could neither identify the sign reversal behavior nor the maxima -- again in agreement with the inhomogeneity scenario which relies on the superposition of AHE contributions with different signs. 

\section{Conclusion}
We conclude that the Hall effect maxima in our SRO films originate from inhomogeneous ferromagnetism. Since inhomogeneity due to deviations from stoichiometry -- including particularly Ru vacancies -- is unavoidable in thin-film deposition of SRO and other ruthenates, we suggest that this scenario should be adopted as the default interpretation of Hall effect maximia in SRO films and heterostructures, and that prior claims of the THE in such structures should be revisited. 

Our results on a well-characterized model compound show in an exemplary fashion how Hall effect maxima can arise from the superposition of positive and negative AHE signals in conjunction with inhomogeneous ferromagnetism. Whereas the presence of two contributions to the AHE with opposite signs is rare for a single phase-pure compound, there are multiple routes towards related situations in heterostructures, multilayers, and composites of different materials and stoichiometries. For instance, AHE contributions of opposite sign can arise through anti-parallel alignment of the magnetization directions of two ferromagnets in a heterostructure, even if the signs of the AHE in the individual compounds are identical. This situation was recently described in heterostructures of elemental ferromagnets \cite{Xu2010}, but deserves careful consideration also for more complex systems where multiple spin systems with non-collinear alignment often arise due to interfacial effects. The sign of the AHE can also change as a function of composition, as recently demonstrated for binary ferromagnetic alloys \cite{Shi2016}. This route towards coexisting positive and negative AHE contributions is particularly relevant for complex oxides -- including manganates -- where compositional variations and associated phase separation are widely observed and difficult to avoid \cite{Moreo1999}. In oxide heterostructures, such variations generically occur as a consequence of charge transfer at interfaces. 

In view of these findings and considerations, we suggest that the widely practiced identification of topological spin textures based on Hall effect maxima is untenable. We note that magnetic force microscopy measurements showing a patchy domain structure, which are often cited in support of the skyrmion interpretation, can be equally well described as small ferromagnetic domains \cite{Wang2018,Meng2019,Vistoli2019}. Discriminating between both pictures requires detailed information about non-collinearity of the magnetization on the atomic scale. Experimental methods that are capable of probing the magnetization in thin-film structures with sufficient sensitivity and resolution include Lorentz microscopy \cite{Yu2010}, coherent x-ray scattering \cite{Li2019}, and diamond quantum sensors \cite{Dovzhenko2018}.


\begin{acknowledgments}
	We thank B. Lemke, B. Stuhlhofer, P. Specht, S. Schmidt for technical supports. 
	We acknowledge E. Goering for the MPMS instrument.
	The project was supported by the European Research Council under Advanced Grant No. 669550 (Com4Com). 
	
	P.A.v.A and Y.E.S. acknowledge support from the European Union's Horizon 2020 research and innovation programme under Grant Agreement No. 823717-ESTEEM3. 
	
	The work at Cornell was supported by the National Science Foundation (Platform for Accelerated Realization, Analysis and Discovery of Interface Materials (PARADIM)) under Cooperative Agreement No. DMR-1539918, the W.M. Keck Foundation, and the Gordon and Betty Moore Foundation's EPiQS Initiative through Grant No. GBMF3850. N.J.S. acknowledges support from the National Science Foundation Graduate Research Fellowship Program under Grant No. DGE-1650441.	
\end{acknowledgments}

\newpage

\vspace{100 mm}
\newpage
$ $
\newpage
\bibliography{apssamp}

\end{document}